\documentclass[notitlepage,superscriptaddress]{revtex4-1}
\usepackage{graphicx,amsmath,amssymb}
\usepackage[usenames]{color}
\usepackage{indentfirst}
\usepackage{float}
\begin{document}
\title{The asymmetric quantum Rabi model in the polaron picture}

\author{Maoxin Liu}
\affiliation{Beijing Computational Science Research Center, Beijing 100084, China}
\author{Zu-Jian Ying}
\affiliation{Beijing Computational Science Research Center,
Beijing 100084, China}
\affiliation{CNR-SPIN, I-84084 Fisciano (Salerno), Italy and Dipartimento di Fisica ``E. R. Caianiello", Universit$\grave{a}$  di Salerno, I-84084 Fisciano (Salerno), Italy}
\author{Jun-Hong An}
\affiliation{Center for Interdisciplinary Studies $\&$ Key Laboratory for
Magnetism and Magnetic Materials of the MoE, Lanzhou University, Lanzhou 730000, China}
\author{Hong-Gang Luo}
\affiliation{Center for Interdisciplinary Studies $\&$ Key Laboratory for
Magnetism and Magnetic Materials of the MoE, Lanzhou University, Lanzhou 730000, China}
\affiliation{Beijing Computational Science Research Center, Beijing 100084, China}
\author{Hai-Qin Lin}
\affiliation{Beijing Computational Science Research Center, Beijing 100084, China}

\begin{abstract}
The concept of the polaron in condensed matter physics has been extended to the Rabi model, where polarons resulting from the coupling between a two-level system and single-mode photons represent two oppositely displaced oscillators. Interestingly, tunneling between these two displaced oscillators can induce an anti-polaron, which has not been systematically explored in the literature, especially in the presence of an asymmetric term. In this paper, we present a systematic analysis of the competition between the polaron and anti-polaron under the interplay of the coupling strength and the asymmetric term. While intuitively the anti-polaron should be secondary owing to its higher potential energy, we find that, under certain conditions, the minor anti-polaron may gain a reversal in the weight over the major polaron. If the asymmetric amplitude $\epsilon$ is smaller than the harmonic frequency $\omega$, such an overweighted anti-polaron can occur beyond a critical value of the coupling strength $g$; if $\epsilon$ is larger, the anti-polaron can even be always overweighted at any $g$. We propose that the explicit occurrence of the overweighted anti-polaron can be monitored by a displacement transition from negative to positive values. This displacement is an experimentally accessible observable, which can be measured by quantum optical methods, such as balanced Homodyne detection.
\end{abstract}

%\date{\today}

\maketitle

\section{Introduction}\label{intro}
The polaron concept \cite{landau} plays a fundamental role in understanding the interaction between electrons and atoms in a solid material and the related transport of the material. In order to effectively screen the charge of an electron moving in the material, the atoms shift from their equilibrium positions, which can be described as a phonon in the language of the second quantization. The electron plus the phonon thus form a quasiparticle -- the polaron, which is a basic entry point to discussing the electron--phonon interaction in many standard textbooks (see, e.g., Ref.\cite{Mahan2000}).

In the past decades, the polaron has found many applications not only in condensed matter physics such as organic semiconductors \cite{deibel1,watanabe}, quantum dots \cite{sauvage}, and high-$T_c$ superconductors \cite{mott,bulaevskii} (one can refer to Ref. \cite{mitra} for an early review, or for a more recent review, see Ref. \cite{devreese}) but also in a variety of other fields such as, for example, Bose--Einstein condensation \cite{cucchietti, bruderer}, nanoscience \cite{snyman}, biophysics \cite{henderson}, polymers \cite{deibel2,tautz,asadi}, and so on. Recently, it has also been extended to study spin coherence \cite{chin1,chin2} in a dissipative quantum system, namely, the spin--boson model \cite{Leggett1987, weissbook}. While these works \cite{chin1,chin2} have focused on the dissipative environment, the physics of polarons in a single-mode optical field have not been
explored systematically. In fact, a two-level atom (qubit) coupled to a single-mode optical field is the simplest system describing the interaction between light and matter, which is a basic model in quantum optics \cite{cavityQED, scully}, quantum information \cite{raimond}, and condensed matter physics \cite{holstein}. This model is well known as the Rabi model \cite{rabi}, which has attracted much attention in the past few years. On the one hand, the exact solutions of this model \cite{Braak2011,Braak2012} and its cousins
\cite{exact_chenqh,prx_xie,Batchelor1,Batchelor2} have been obtained only recently, which marks an important step toward the understanding of the mathematical structure of this model by proving its integrability \cite{Braak2011, Solano2011,Cibils,Stepanov,moroz2016,moroz2014}. On the other hand, the experimental progress \cite{Romero2012,Restrepo2014,Crespi2012,circuit1,circuit2,Niemczyk2010,Forn-Diaz2010} toward significant enhancement of the light--matter coupling strength raises the request to consider a full quantum Rabi model in order to explain experimental observations \cite{Niemczyk2010, Forn-Diaz2010}.

However, since numerically solving for the transcendental function zeros in the exact solution is not an easy task, understanding the model from some simple viewpoints is still needed. To explore the physical properties of the Rabi model in some intuitive ways, many approximation methods have been developed; these include, for example, the adiabatic approximation \cite{adiabatic}, the generalized rotating-wave approximation \cite{grwa,grwa_bias}, and the generalized variational method \cite{GVMground}. A systematic comparison study of these methods has been given in Ref. \cite{Liu2015}. It was shown that all these methods are unable to capture the correct physics in the intermediate coupling strengths, especially in the small $\omega$ (harmonic oscillator frequency) limit. Apart from the double precision problem in eigenstate calculations\cite{Braak2012,moroz2014} and the careful attention one might need to pay to the criterion and parameter regime for quantum integrability \cite{Batchelor1,Cibils,Stepanov,moroz2016}, it also might be worthwhile to recognize that the basic energy scales involved in the Rabi model compete with each other in a subtle way \cite{Ying2015}. All these factors might contribute to the fact that the physics of the quantum Rabi model still has not been fully explored. Thus, one should face the challenge of using an intuitive way to understand the Rabi model in a general case, in which extensive physics phenomena might be involved \cite{ashhab1,ashhab2,variation,critical1,critical2,critical3}. In our previous work \cite{Ying2015}, we found that the deformed polaron picture is a good starting point to study the Rabi model in the whole parameter regime, and furthermore, a ground state phase diagram of the Rabi model has been extracted and some underlying physics unveiled. Essentially, this diagram consists of two regimes, namely, the bipolaron regime in the strong coupling and the quadpolaron regime in the weak coupling, with a deep nature of symmetry breaking in their transition or crossover. Interestingly, in the latter regime, an overweighted anti-polaron region has been found. While the basic physics of the Rabi model has been clarified \cite{Ying2015}, much less attention has been paid to the asymmetric case, in which the parity symmetry is further broken. In fact, from the experimental viewpoint, the asymmetric case is more usual. Therefore, it is important to study how the finite asymmetric term affects the basic properties of the Rabi model.

The presence of the asymmetric term evidently increases the difficulty of solving the model, owing to parity breaking; therefore, we first numerically solve the asymmetric Rabi model for its ground state wave function. It is found that the asymmetric term tends to trigger the appearance of the so-called overweighted anti-polaron. Based on the polaron picture, we analyze the competition of a variety of energy scales involved in the asymmetric Rabi model, and clarify how an overweighted anti-polaron explicitly appears. If the strength of the asymmetric term is less than the harmonic oscillator frequency, it is found that the overweighted anti-polaron can occur at a certain coupling strength $g_0$, at which the entanglement entropy drops rapidly to a vanishingly small value and the sign of the displacement of the up-spin component changes from negative to positive. If the strength of the asymmetric term is greater than the harmonic oscillator frequency, the negatively displaced components are totally suppressed and the anti-polaron in the up-spin component is always overweighted. These results could be tested by measuring the displacement of the harmonic oscillator in the up- or down-spin components via the balanced Homodyne detection method \cite{homodyne} in quantum optics.

The paper is organized as follows. In Sec. \ref{model}, the Rabi model with the asymmetric term is introduced and formulated in position representation. In Sec. \ref{polaron}, based on the polaron picture, we numerically solve the symmetric and asymmetric Rabi model and obtain the ground state wave function for different coupling strengths. The occurrence of the overweighted anti-polaron and its mechanism are analyzed. In Sec. \ref{observable}, we calculate the displacements of the up- and down-spin components, which provide an experimentally accessible physical quantity to detect the overweighted anti-polaron. In Sec. \ref{variational}, by a variational discussion, the connection between the wave function based on the polaron/anti-polaron picture and the approximate methods widely used in the literature is clarified, which facilitates an analytical derivation of the critical asymmetry strength. Finally, Sec. \ref{section_con} is devoted to a brief summary.

\section{Model}\label{model}
The Hamiltonian of the asymmetric quantum Rabi model \cite{ec,arabi1,arabi2,arabi3} ($\hbar=1$) is
\begin{equation}\label{arabi}
H_{A} = \omega a^{\dagger}a+g\sigma_x(a+a^\dagger)+\Delta \sigma_z+\epsilon '\sigma_x,
\end{equation}
where $a$ ($a^{\dag }$) is the annihilation (creation) operator of
the bosonic field with frequency $\omega $, $\sigma_i (i = x,z)$
is the Pauli matrix of the qubit, $2\Delta$ is the energy splitting of the qubit,
$\epsilon'$ is a strength breaking the parity symmetry of the system, and $g$, which must be positive, is the coupling strength.
One can transform Eq.\eqref{arabi} to the following equivalent Hamiltonian by the spin rotation $H=e^{i\frac{\pi}{4}\sigma_y}H_Ae^{-i\frac{\pi}{4}\sigma_y}$ as well as the parameter replacements $\epsilon'=\frac{\epsilon}{2}$ and $\Delta=-\frac{\Omega}{2}$:

\begin{equation}\label{rabi}
H = \omega a^{\dagger}a+\frac{\Omega }{2}\sigma_x+\frac{\epsilon}{2}\sigma_z+g\sigma_z(a+a^\dagger),
\end{equation}
which has the same notation as the spin--boson model\cite{Leggett1987,weissbook} on a $\sigma_z$ basis. Thus, $\Omega$, which is positive and taken as the unit of energy unless otherwise specifically stated, determines the tunneling rate between the two eigenstates of $\sigma_z$. $\epsilon$ is the asymmetry strength. In this work, the coupling strength $g$ will be scaled by a critical value $g_c = \frac{\sqrt{\omega\Omega}}{2}$ \cite{ashhab1,ashhab2,rabiQPT,zpb1984}, where a superradiance phase transition occurs \cite{sr,rabiQPT,JCQPT}.

It should be mentioned that the Rabi model has wide relevance and applications, ranging from quantum optics\cite{cavityQED}, superconducting circuit quantum electrodynamics\cite{circuit1, circuit2, Niemczyk2010,Irish2014, Forn-Diaz2010, grwa_bias, Albert2012, Albert2011} to condensed matter physics\cite{holstein}. The notation may be different in different communities, as pointed out by Irish\cite{grwa}. Hereafter, for the convenience of our analysis, we shall discuss on the basis of Eq.\eqref{rabi}, whose symmetric part has also been widely used in the literature\cite{grwa, adiabatic, Irish2014, Forn-Diaz2010, grwa_bias, Albert2012, Albert2011, circuit1, circuit2, Niemczyk2010, critical1, chin1, chin2, ashhab1, ashhab2, GVMground, Liu2015, Liu2013}.

In terms of the coordinate operator
$x=\sqrt{\frac{1}{2\omega}}(a+a^{\dag})$ and momentum operator
$p=i\sqrt{{\omega\over 2}}(a^{\dag}-a)$, Eq. (\ref{rabi}) can be
reformulated as
\begin{equation}\label{xp_dis}
H=H_{\text{DO}}+\frac{\Omega }{2}\sigma_x+\frac{\epsilon}{2}\sigma_z -\frac{g^2}{\omega},
\end{equation}
where $H_{\text{DO}}=\frac{1}{2}[\omega^2(x+\sigma_z x_0)^2+p^2]$,
with $x_0= \sqrt{2g^2/\omega^3}$, describes a displaced harmonic
oscillator with the displacement direction conditioned by
the qubit state. One sees that the displacement is driven by
the interaction and simultaneously also affected by the mode
frequency $\omega$. The eigenstate of Eq. (\ref{xp_dis}) can take
a general form
\begin{equation}\label{general_wave}
|\Psi\rangle=|+_z\rangle\phi_+(x) + |-_z\rangle\phi_-(x),
\end{equation}
where $|\pm_z\rangle$ are the eigenstates of $\sigma_z$ and here $\pm$ denotes the spin orientation, and $\phi_{\pm}(x)$, the corresponding oscillator states, satisfy $\phi_-^2(x)+\phi_+^2(x)=1$.

The eigenstates are determined by the following Schr\"odinger
equations coupled for up- and down-spins,
\begin{equation}
\left[ \frac{\hat p^2}2+\frac{\omega ^2}2\left( \hat x\pm
x_0\right) ^2\pm \frac \epsilon 2\right] \phi _{\pm }+\frac
\Omega 2\phi _{\mp }=E\phi _{\pm },
\end{equation}
which will be solved numerically in the present work by an expansion on the basis set of the standard quantum harmonic oscillator \cite{Crespi2012} with a truncation judged via the distribution weight on a basis, which is cut off if below a sufficiently small value. Here, we employ the basis $|\pm_z\rangle \otimes |n\rangle$, where $|n\rangle$ is the Fock state, and the preserved Hilbert space is up to 6000-dimensional, so that the distribution weight is smaller than $10^{-13}$. We believe that false convergence \cite{krabi} is prevented by considering this large preserved Hilbert space. A deep discussion on the convergency of the numerical method for the Rabi model can be seen in Ref. \cite{rabiconverge}.  One sees that the $ \Omega $ term plays a role of spin flipping; it is also referred to as the tunneling term, owing to an effective barrier between the two wells of the left and right displaced harmonic potential in the single-particle description\cite{Irish2014,Ying2015}.

\section{Overweighted anti-polaron and its mechanism in the Rabi model}\label{polaron}
\subsection{The polaron and anti-polaron in the symmetric case}
To provide a reference for the asymmetric case, we first illustrate the symmetric case.
For the symmetric case, i.e., $\epsilon=0$, one can find that the
parity operator $\Pi=\sigma_x(-1)^{a^{\dag}a}$ is conserved
because $[\Pi, H] = 0$. Thus, Eq. (\ref{general_wave}) is also the
eigenstate of $\Pi$, which results in
\begin{equation}
\phi_+(x)=\pi\phi_-(-x),
\end{equation}
where $\pi=\pm 1$, and $+1$ and $-1$ represent even
and odd parities, respectively. In this work, we only discuss the
ground state, which is in odd parity.

In the absence of the tunneling term $\frac{\Omega}{2}\sigma_x$, the ground state of the Hamiltonian (\ref{xp_dis}) consists of two polarized states surrounded by photons that are induced by displacement from the central origin position. As is known, the ground state of the harmonic oscillator is a photon vacuum, but a displaced state is accompanied by photons. This resembles the polaron in condensed matter, where the electron is surrounded with photons that are induced by deviation of atoms from their equilibrium positions. Thus, these right ($x_0$) and left ($-x_0$) displaced states can be considered as polarons\cite{chin1,chin2,Ying2015}.
The polaron located
at $x_0$ is bounded to $|-_z\rangle$ and that located at
$-x_0$ is bounded to $|+_z\rangle$. When the $\Omega$ term is
gradually increased, the displacement and the tunneling between
two polarization states start to compete. On the one hand,
the polarons prefer to reside at the positions of $\pm x_0$
in order to avoid the cost from higher harmonic oscillator
potential in deviating away. On the other hand, the presence
of the tunneling requires that these two polarons should be as
close as possible to increase the effective tunneling. As a
consequence of this competition, anti-polarization
components arise; namely, an anti-polaron with negative
displacement is bounded to $|-_z\rangle$ and one with a
positive displacement is bounded to $|+_z\rangle$, in the
opposite displacement directions of the polarons. In this case,
the harmonic oscillator wave function components in Eq.
(\ref{general_wave}) in both the up- and down-spins can be divided
into two parts:
\begin{equation}\label{pa_wave}
\phi_\pm(x)=\phi_\pm^{\rm P}(x)+\phi_\pm^{\rm A}(x),
\end{equation}
where the subscript $\pm$ labels the spin orientation and the
superscripts ${\rm P}$ and ${\rm A}$ denote the polaron and
anti-polaron, respectively. For simplicity, here, we have
incorporated the weights of the polaron and the
anti-polaron\cite{Ying2015} into $\phi_\pm^{\rm A,P}$ so that the
height of the wavepacket peaks will directly reflect the weights
of the separated polaron and anti-polaron. Subject to the
constraint of parity symmetry, the harmonic oscillator
wave function components satisfy the following relations:
\begin{equation}\label{symmetry}
\begin{split}
\phi_-^{\rm{P}}(x)&=-\phi_+^{\rm{P}}(-x),\\
\phi_-^{\rm{A}}(x)&=-\phi_+^{\rm{A}}(-x).
\end{split}
\end{equation}

\begin{figure}
\includegraphics[width=0.8\columnwidth]{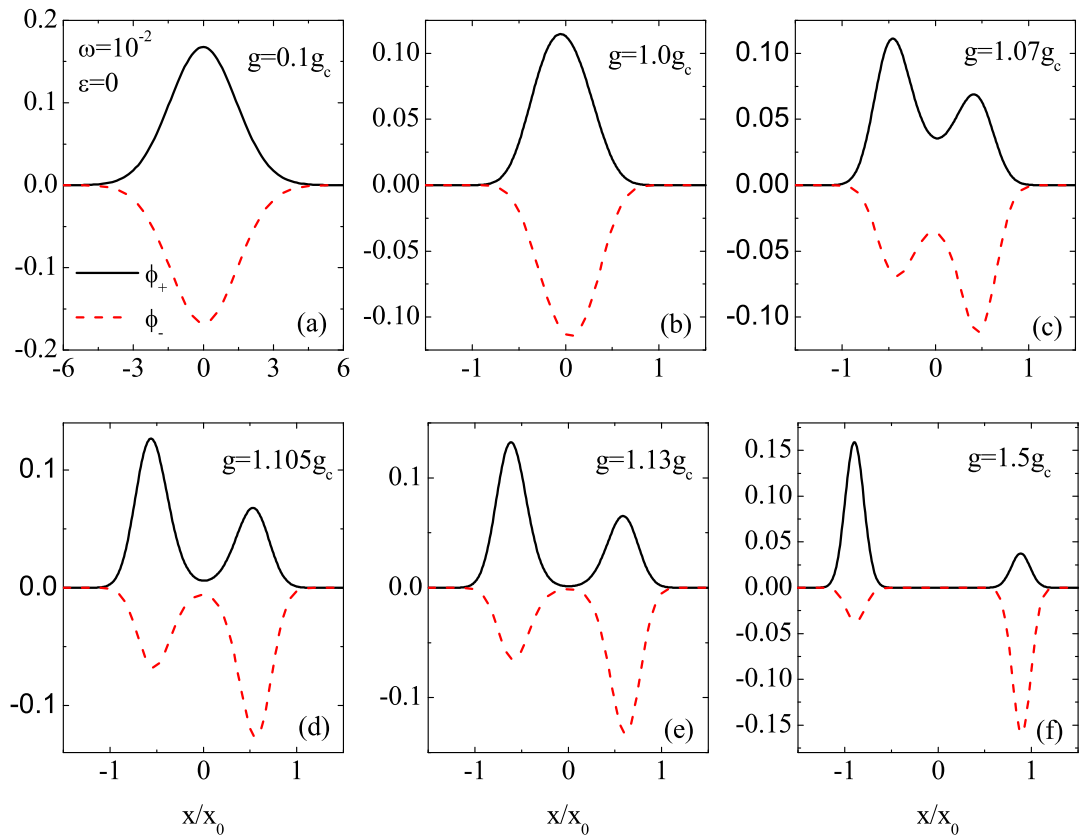}
\caption{(Color online) Ground state wave function of the symmetric Rabi model with $\omega=0.01$, $\epsilon=0$, and $g=0.1g_c$ in (a), $g=1.0g_c$ in (b), $g=1.07g_c$ in (c), $g=1.105g_c$ in (d), $g=1.13g_c$ in (e), and $g=1.5g_c$ in (f).  }
\label{wave_un}
\end{figure}

To illustrate the profiles of the polaron and anti-polaron,
the numerical result of the ground state wave function with frequency
$\omega = 0.01$ is shown in Fig. \ref{wave_un} for different
coupling strengths (in the symmetric model).
Increasing the coupling strengths, two characteristic features have been
observed. One is the separation of the polaron and anti-polaron
peaks. At sufficiently weak coupling, as shown in Fig. \ref{wave_un}
(a), the tunneling dominates over the polaron displacements, and as a result, the polaron and anti-polaron
overlap so much that, finally, only one total peak is observed
around the center position. As the coupling strength increases
to $g_c$, the peaks begin to deform and shift away from the
center position, as seen in Fig. \ref{wave_un} (b). The change is
more obvious by further increasing the coupling strengths, as
shown in Fig. \ref{wave_un} (c)--(f). The two separated peaks gradually shift
toward the position of $\pm x_0$, where the coupling
dominates over the tunneling. The other feature lies in the
weights of the polaron and anti-polaron. Although the weight
variation in the weak and intermediate couplings is
diversified\cite{Ying2015}, after full separation the weight of
the anti-polaron becomes smaller and smaller. This is
understandable, since the anti-polaron needs to overcome the
increasing cost of higher and higher harmonic oscillator
potentials. One notes on these characteristic features and their
transitions can be well described by the energy scale $g_c$
defined above, which has been shown to be valid only in
small-$\omega$ limit \cite{Ying2015}. In this case, these
transitions show phase-transition-like behavior
\cite{ashhab1,ashhab2,variation,critical1,critical2,critical3}.
It has been pointed out that in the small-$\omega$ limit, there is indeed a quantum phase transition between the normal and superradiance phases\cite{sr,rabiQPT}. From Fig. \ref{wave_un} (b) and (c), where the coupling is around the critical coupling $g_c$, we see that the polaron and anti-polaron start to displace and separate in a narrow interval, which indicates a phase transition. We can also find that the so-called superradiance phase exhibits displacement and separation between the polaron and anti-polaron.
For a larger $\omega$, the definition of $g_c$ here is no
longer a good energy scale \cite{Ying2015}, and these transitions
cross over, as will be shown later. In the following we
consider the physics with finite asymmetry strength.

\subsection{Overweighted anti-polaron in asymmetric model}\label{neq}

\begin{figure}
\includegraphics[width=0.8\columnwidth]{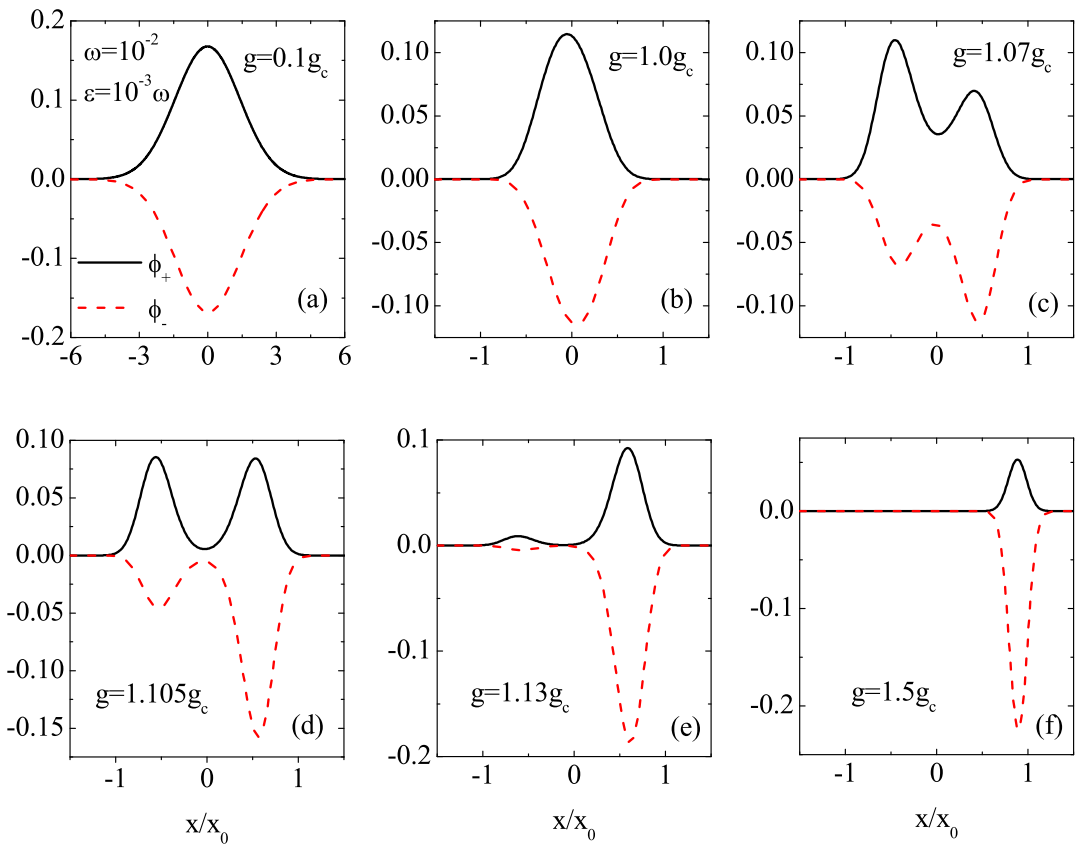}
\caption{(Color online) Ground state wave function of the asymmetric Rabi model with $\omega=0.01$ and $\epsilon=10^{-3}\omega$. The coupling strengths are the same as those in Fig. \ref{wave_un}.}
\label{wave}
\end{figure}

In the asymmetric model, $\epsilon\neq 0$, the parity symmetry is
broken; as a result, Eqs. \eqref{symmetry} are no
longer valid. As a comparison to the symmetric model, the
numerical results for the ground state wave function are
presented in Fig. \ref{wave} at the same frequency as in Fig.
\ref{wave_un}, $\omega = 0.01$, but in the presence of a finite
asymmetry strength $\epsilon = 10^{-3}\omega$. In the weak coupling regime ($g
\lesssim g_c$), the features of the wave functions are similar to
those in the symmetric model. However, as $g > g_c$, the
characteristic behaviors of the wave functions are dramatically
different. First, the weights of the up- and down-spin components
are different. Second, the polaron and anti-polaron for the
two spin components behave completely differently. For the first
point, it is not unexpected, since the finite asymmetry strength is more favorable for the down-spin component. However, the
second point is counter-intuitive and nontrivial. For the
down-spin component, the weight of the anti-polaron (right
wave-packets in dashed lines in Fig. \ref{wave}(c)--(f)) decays
sooner in comparison with that of the symmetric version.
More dramatically, the weight of the polaron (right wave-packets
in solid lines) in the up-spin component decays even more
rapidly, and these  disappearing weights move totally to the
polaron in the down-spin component (left wave-packets in
dashed lines) with increasing coupling strength.

In this process, the weight of the anti-polaron in the up-spin component
(left wave-packets in solid lines) becomes larger than that
of the polaron (right wave-packets in solid lines), as shown
in Fig. \ref{wave} (e) and (f). This is the situation of the
overweighted anti-polaron.
Therefore, the appearance of the
overweighted anti-polaron in the up-spin component is here due to
the interplay between the tunneling, coupling of light and
matter, and finite asymmetry strength. The overweighted anti-polaron also
occurs in the symmetric Rabi model in an implicit way, as uncovered in Ref.
\cite{Ying2015}. However, there, the overweighted anti-polaron appears symmetrically for both spin directions with a different mechanism in
the subtle competition between the tunneling, coupling, as
well as the harmonic frequency, and it only occurs in certain
regions. Here, a dominant role is played by the finite asymmetry strength: all weights of
the negative displacement and the anti-polaron in the up-spin
component will vanish as the coupling strength increases. Thus,
in the finite asymmetry strength case, the overweighted anti-polaron in the
up-spin component must occur beyond a certain coupling
strength. This is the central result of the present work.

In the following we further analyze the physics of the
model from the viewpoint of energy competition. In the asymmetric Rabi
model, there exist four energy scales, namely, the harmonic
frequency $\omega$, tunneling rate $\Omega$ (set to unity), coupling
strength $g$, and asymmetry strength $\epsilon$. First, the
competition between the effective tunneling and coupling strength, which drives the displacement of the potential at a given frequency as indicated by $x_0$, determines whether the polaron and anti-polaron components get separated, and this
phase-transition-like behavior is described by the parameter $g_c$
in the small-$\omega$ case. Second, once the polaron and anti-polaron components become separated, tunneling between the left and right sides, which is unfavorable for the asymmetry strength, is significantly weakened. The tunneling, remaining on same sides, becomes more compatible with the asymmetry strength and these may even support each other. Thus, the interplay between asymmetry strength and the remaining effective tunneling, in competition with the potential energy cost for increasing the anti-polaron weight, determines whether the negative displacement components are suppressed and whether the overweighted anti-polaron appears. This changeover turns out to be the second phase-transition-like behavior.

To further understand the competitions, we combine the analysis with an explicit situation in the example
we illustrated. For the parameters we have used in
Fig. \ref{wave}, the energy scales involved are $\Omega \gg
\omega \gg \epsilon$ and $g_c = 0.05\Omega$. Therefore, for a
sufficiently small coupling strength $g \lesssim g_c$, the
tunneling energy (characterized by $\Omega$) is dominant. In this
situation, the oscillator displacement is totally suppressed,
not to mention the even weaker asymmetry strength, which is completely
overwhelmed by the tunneling energy. In this case, the asymmetry strength
plays a minor role in determining the ground state. However, as
the coupling strength increases, e.g., $g > g_c$, the situation
changes dramatically. In fact, the polaron and anti-polaron
will shift away from the central position, owing to the increasing
coupling. As a consequence of the displacements, the effective
tunneling between the polaron and polaron, polaron and anti-polaron,
as well as anti-polaron and  anti-polaron decays exponentially. Therefore, when the effective tunneling decays to a
small value comparable to the asymmetry strength, the asymmetry strength comes to play a
considerable role. In such a situation, the asymmetry strength interplays with
the tunneling, since moving the weights of the left-hand-side polaron
and anti-polaron to the right-hand-side polaron is
favorable both for the asymmetry strength and tunneling energies. This mutually
beneficial interplay quickly wins its competition at the cost
of the coupling-and-frequency-dependent potential energy in weight
moving and thus, all weights in the up-spin components and those of
the anti-polaron of the down-spin components transfer
rapidly to the polaron in the down-spin components. As a
result, an overweighted  anti-polaron in the spin-up component occurs.
This also behaves like a phase transition, as seen from the
entanglement entropy calculations for different asymmetry strengths.

\begin{figure}
\includegraphics[width=0.8\columnwidth]{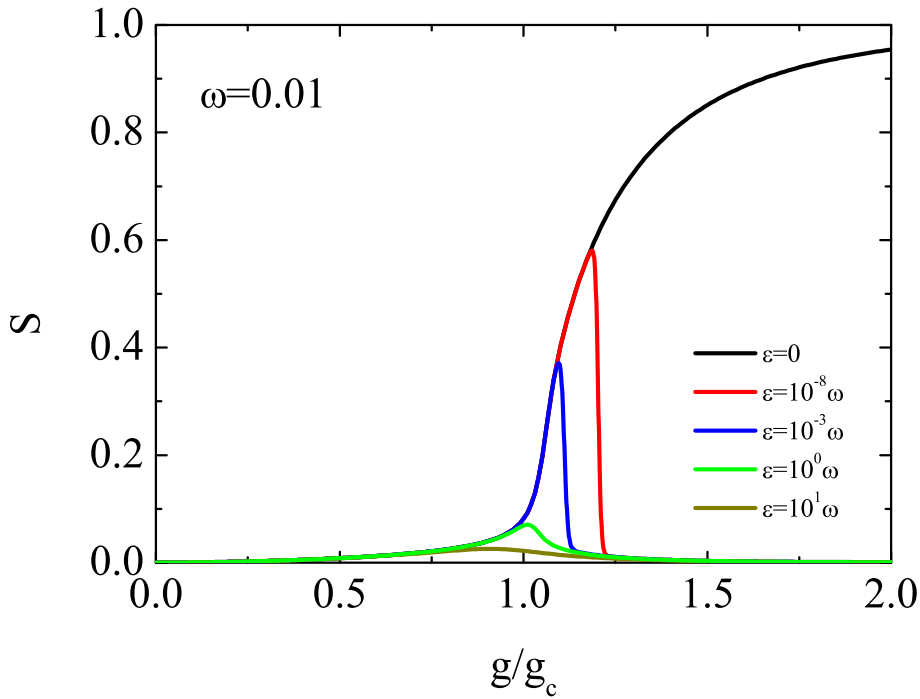}
\caption{(Color online) Entanglement entropy as a function of coupling strength with $\omega=0.01$ for different asymmetry strengths: $\epsilon=0\omega$ (black), $\epsilon=10^{-8}\omega$ (red), $\epsilon=10^{-3}\omega$ (blue), $\epsilon=10^{0}\omega$ (green), and $\epsilon=10^{1}\omega$ (dark yellow).}
\label{001s}
\end{figure}

The entanglement entropy can be quantified by von Neumann entropy:
\begin{equation}
S=-\text{Tr}_\text{q}\{\rho_q \log_2 \rho_q\},
\end{equation}
where $\rho_q=\text{Tr}_{\text{osc}}\{|\Phi\rangle\langle\Phi|\}$
is the reduced density matrix of the qubit. The result is
presented in Fig. \ref{001s}.
It has been noticed that the
behaviors of entanglement entropy in the symmetric and asymmetric models are quite
different \cite{ashhab1}. Here, based on the polaron and anti-polaron concept,
we present a physical picture to understand the behavior of entanglement.
As seen from Fig. \ref{001s}, the entanglement entropy
has two phase-transition-like points; one is near $g_c$, marking
the separation of the polaron and anti-polaron, and there the entropy
increases dramatically. In the symmetric Rabi model, the entropy
saturates in the strong coupling limit. However, in the asymmetric model, after the entropy increases to certain value, it drops
rapidly to a vanishing small value, marking another
phase-transition-like behavior.
This is due to the asymmetry-induced anti-polaron, which  efficiently decreases the entanglement.
When the asymmetry strength increases, the latter
phase-transition-like behavior happens earlier, but
the asymmetry finally becomes dominant and the phase-transition-like behaviors
cross over.

\section{Observable effect of the anti-polaron}\label{observable}

So far, we have revealed the appearance of an overweighted
anti-polaron around an additional transition-like behavior induced
by the asymmetric term. Now, a natural question arises: whether
the overweighted anti-polaron can be observed in an
experiment. To answer this question, let us consider a physical
quantity, $X_{\pm}$, which is
defined as an average of the oscillator state of either the
up-spin or down-spin, namely,
\begin{equation}
X_{\pm}=\langle a^{\dag}+a\rangle_{\pm}=\langle
\phi_{\pm}(x)|(a^{\dag}+a)|\phi_{\pm}(x)\rangle.
\end{equation}
$X_{\pm}=\langle a^{\dag}+a\rangle_{\pm}$ describes the
displacement of the oscillator state $\phi_{\pm}(x)$.

\begin{figure}
\includegraphics[width=0.6\columnwidth]{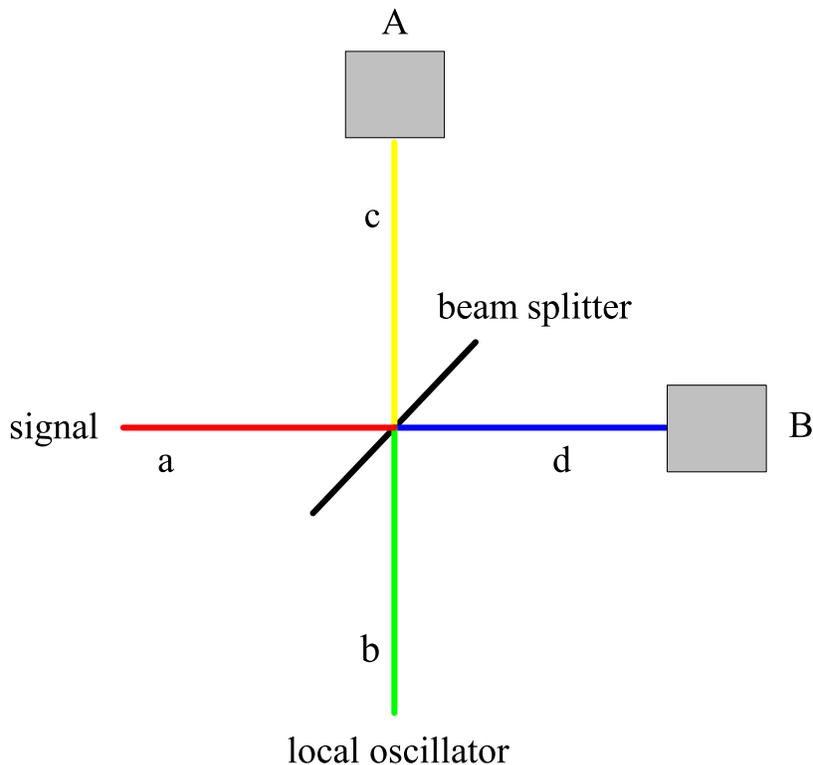}
\caption{(Color online) Illustration of the measurement of $a+a^{\dag}$ by balanced Homodyne detection.}
\label{BHD}
\end{figure}

The prerequisite of a measurement on $X_{\pm}$ is to obtain the oscillator in either $\phi_+(x)$ or $\phi_-(x)$. For this purpose, we follow the scheme in Ref. \cite{ashhab1}. First, we measure the spin $\sigma_z$. The state after the measurement collapses to $\phi_+(x)$ (or $\phi_-(x)$) if the measurement result is $+1$ (or $-1$). Note that $\sigma_z$ should be $\sigma_x$ when we transform back to Eq. \eqref{arabi}. Second, we measure $X_{\pm}$ by the position operator $a+a^{\dag}$ by the so-called balanced Homodyne detection \cite{homodyne} method. A measurement scheme of $a+a^{\dag}$ is briefly illustrated in Fig.~\ref{BHD}. The oscillator in state $\phi_{\pm}$ is input via port ``a". A modulating local oscillator in a coherent state $|\alpha\rangle$ with $\alpha=|\alpha|e^{-i\frac{\pi}{2}}$ is input via port b. The beam splitter is ideally half-transparent. Thus, the detectors ``A" and ``B" can measure both outcomes $c=\frac{1}{\sqrt{2}}(a+ib)$ and $d=\frac{1}{\sqrt{2}}(ia+b)$, respectively. The counting imbalance between the two detectors obtains the expectation value of the number difference operator $M=d^{\dag}d-c^{\dag}c=-i(a^{\dag}b-b^{\dag}a)$. Thus, we can obtain $\langle a+a^{\dag} \rangle= -\frac{M}{|\alpha|}$.

\begin{figure}
\includegraphics[width=0.8\columnwidth]{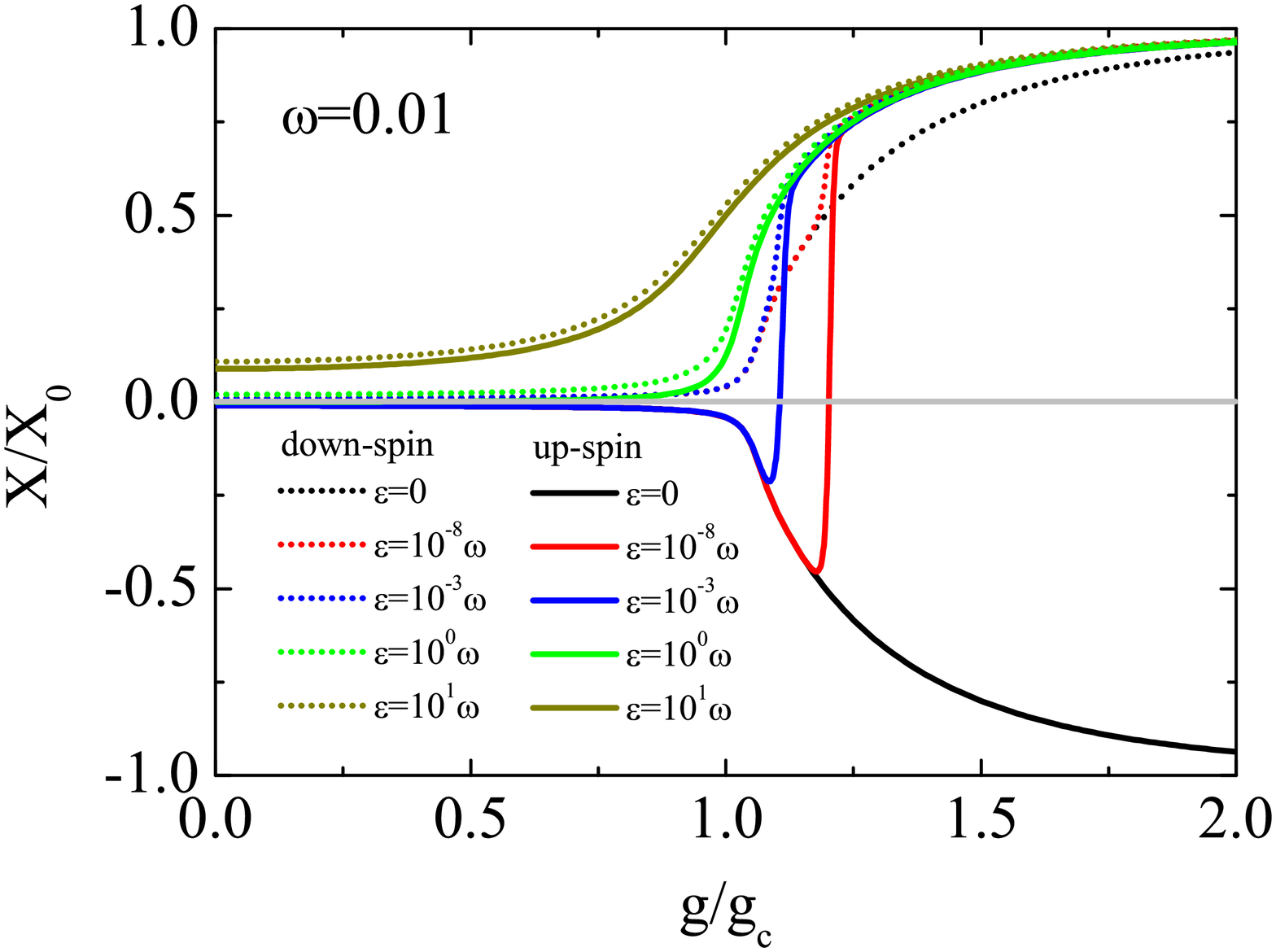}
\caption{(Color online) Oscillator displacement of both spin-up and spin-down as a function of coupling strength with $\omega=0.01$ for different asymmetry strengths: $\epsilon=0\omega$ (black), $\epsilon=10^{-8}\omega$ (red), $\epsilon=10^{-3}\omega$ (blue), $\epsilon=10^{0}\omega$ (green), and $\epsilon=10^{1}\omega$ (dark yellow).}
\label{001x}
\end{figure}

\begin{figure}
\includegraphics[width=0.8\columnwidth]{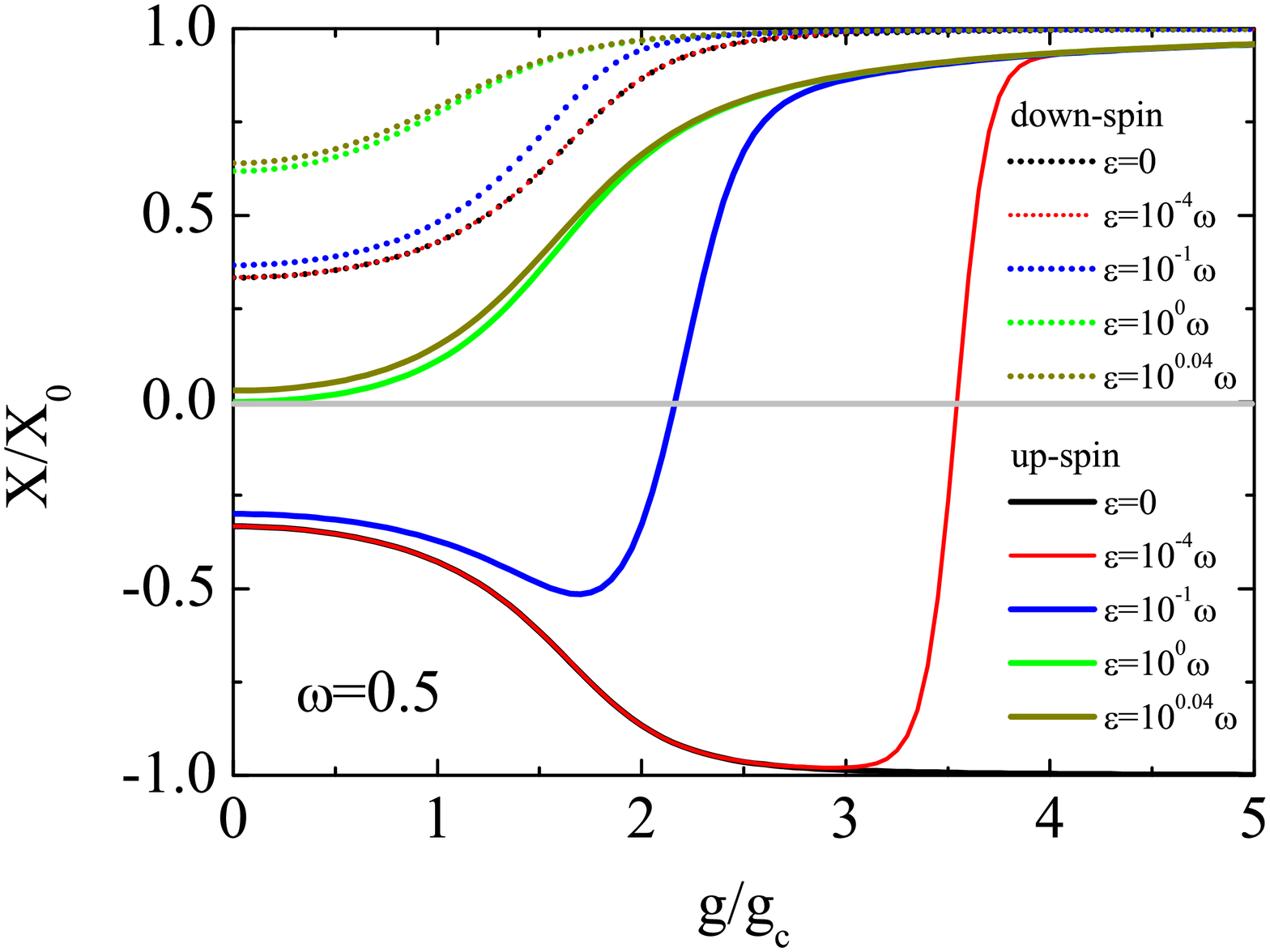}
\caption{(Color online) Oscillator displacement of both spin-up and spin-down as a function of coupling strength with $\omega=0.5$ for different asymmetry strengths: $\epsilon=0\omega$ (black), $\epsilon=10^{-4}\omega$ (red), $\epsilon=10^{-1}\omega$ (blue), $\epsilon=10^{0}\omega$ (green), and $\epsilon=10^{0.04}\omega$ (dark yellow). }
\label{05x}
\end{figure}
\begin{figure}
\includegraphics[width=0.8\columnwidth]{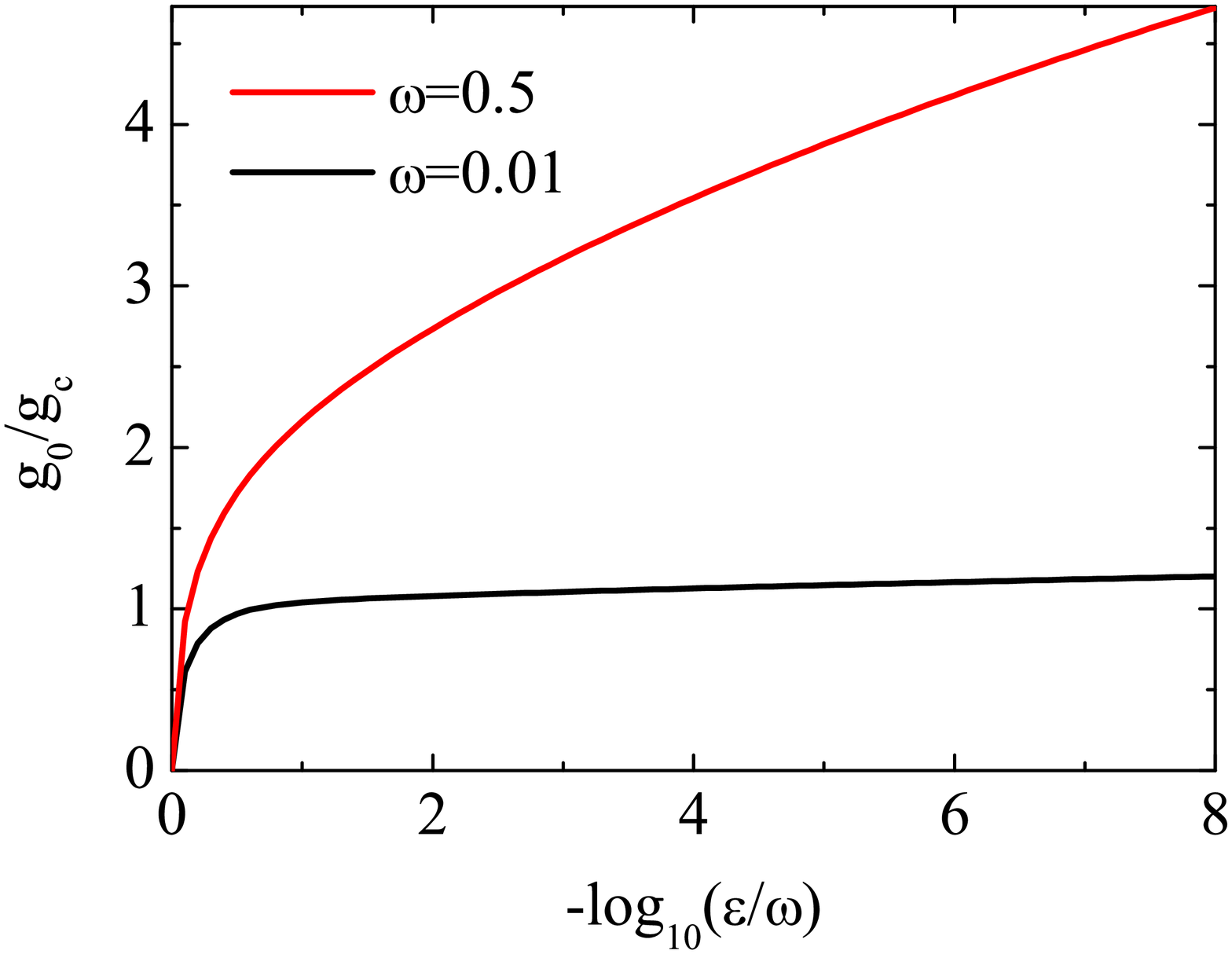}
\caption{(Color online) Displacement transition point $g_0$ as a function of asymmetry strength $\epsilon$ with $\omega=0.5$ (red) and $\omega=0.01$ (black).}
\label{g0}
\end{figure}

Figures \ref{001x} and \ref{05x} show $X_{\pm}$ as a function of coupling strength $g$
for different asymmetry strengths with $\omega=0.01$ and $0.5$, respectively.
We first discuss the small-$\omega$ case, i.e., that shown in Fig. \ref{001x}.
The wave functions shown in Fig. \ref{wave} correspond
to the set of parameters denoted by solid and dotted blue
lines ($\epsilon=10^{-3}\omega$). The solid and dotted black
lines denote the symmetric case. As one sees, in the small
asymmetry strength cases, the harmonic oscillators do not displace far away
from the central position up to $g\sim g_c$. Beyond $g_c$, the
displacements vary rapidly but in different ways, depending
on the absence or presence of the asymmetric term. In the symmetric case, the
displacement amplitude grows quickly and then approaches the
saturation value $X_0=2g/\omega$, with the spin-up and spin-down components being
symmetric. This symmetry in the evolution is broken by finite
asymmetry strength. A striking change can be observed for the spin-up component,
which suddenly stops growing in the negative direction and turns abruptly to become
positive at a certain coupling strength $g_0$. A larger asymmetry strength will
have a smaller sign reversing point $g_0$, whose overall
dependence on the finite asymmetry strength value and the harmonic frequency is
shown in Fig. \ref{g0}. The change at $g_0$ is quite sudden
and similar to a phase-transition, which is consistent with the dropping behavior of the entanglement entropy. Moreover,
when we further increase the asymmetry strength, we find that both
displacements for the up- and down-spin components become
positive throughout the evolution with respect to the
coupling strength, as in the $\epsilon=10^0\omega$, $10^1\omega$
cases in Fig. \ref{001x}. In such strong asymmetry strength cases, the
anti-polaron of the spin-up component is always overweighted.

For a large harmonic frequency, $\omega = 0.5$, the
transition-like behaviors become relatively smooth,
especially that around $g_c$. However, the changeover at
$g_0$ is still quite sharp, as shown in Fig. \ref{05x}, which
indicates that it is easy to observe this transition of
displacement from negative to positive value in a wide parameter
regime.

A further overview on the above results, or a closer
observation of Fig. \ref{g0}, will unveil another critical point in
the asymmetric strength $\epsilon$. In fact, in Fig. \ref{g0}, it is
shown that only for $\epsilon < \omega$, one can always observe this
transition of overweighted anti-polaron for the up-spin component.
The situation becomes more particular in the regime $\epsilon > \omega$. This is the case in which both
displacements for the up- and down-spin components are positive
and the anti-polaron (if it survives) is always overweighted,
as mentioned in the strong asymmetry strength cases at $\omega=0.01$.
Therefore, there exists a critical asymmetry strength $\epsilon_c$ which
determines whether the oscillator displacement transition from
negative to positive value in spin-up component will
occur. Our numerical calculations indicate that:
\begin{equation}
\epsilon_c=\omega.
\end{equation}
In the next section, we provide a variational method to discuss
the result analytically.

\section{Analytic derivation for $\epsilon_c$ by variational method} \label{variational}

 Before deriving the analytic expression for $\epsilon_c$, it might be worthwhile to see the implications that the previously discussed polaron and anti-polaron provide for variational methods.
Taking the polaron concept as a starting point, one can build a connection between the wave function profiles obtained in the present work and the approximate methods widely used in the literature\cite{adiabatic,grwa,grwa_bias,GVMground}. Let us first neglect the anti-polaron part in the wave function; thus, the trial wave function $|\Psi\rangle$ can be assumed to be the superposition of the coherent states with opposite displacements and spins
\begin{equation}\label{trial}
|\Psi\rangle =\cos \frac{\theta}{2}|+_z\rangle|-\lambda\rangle+\sin \frac{\theta}{2}|-_z\rangle|\lambda\rangle,
\end{equation}
where $\theta$ is a variational parameter and $|\lambda\rangle=e^{\lambda(a^{\dagger}-a)}|0\rangle$ is the coherent state with another variational parameter $\lambda$. Then, the energy expression can be obtained as
\begin{equation}
E = \omega\lambda^2-2g\lambda+\frac{1}{2}(\Omega e^{-2\lambda^2}\sin\theta+\epsilon\cos\theta).
\end{equation}
Minimizing the energy with respect to $\theta$ leads to $\tan\theta=\Omega e^{-2\lambda^2}/\epsilon$ and the ground-state energy becomes:
\begin{equation}
E_{\text{G}}(\lambda)\equiv E_\text{min} = \omega\lambda^2-2g\lambda-\frac{1}{2}\sqrt{\Omega^2 e^{-4\lambda^2}+\epsilon^2}.
\end{equation}
If $\lambda$ is fixed to be $\lambda=g/\omega$, those results discussed in Refs. \cite{adiabatic,grwa,grwa_bias} are reproduced.
Furthermore, if $\lambda$ is determined by minimization of the energy $E_{\text{G}}$ with respect to $\lambda$, namely, $\partial  E_{\text{G}}/\partial \lambda=0$,
then the result in Ref. \cite{GVMground, Liu2015} is recovered.

As also mentioned in the introduction, these approximate methods cannot capture the correct result in the intermediate coupling regime owing to the competition between different energy scales, which is comparable in this coupling regime. In this case, the contribution of the anti-polaron component becomes considerable and should be taken into account. Thus, in the spirit of Eq.(\ref{pa_wave}), the trial wave function in Eq. (\ref{trial}) should be modified into a more general form:
\begin{equation}\label{anatz2}
\begin{split}
|\Psi\rangle=&C^-|-_z\rangle(\alpha^-_{\text{P}}|\lambda^-_{\text{P}}\rangle+\alpha^-_{\text{A}}|\lambda^-_{\text{A}}\rangle)\\
&+C^+|+_z\rangle(\alpha^+_{\text{P}}|\lambda^+_{\text{P}}\rangle+\alpha^+_{\text{A}}|\lambda^+_{\text{A}}\rangle),
\end{split}
\end{equation}
The parameters $C^{\pm}$, $\alpha^{\pm}_{\text{P},\text{A}}$, and $\lambda^{\pm}_{\text{P},\text{A}}$ can be determined by minimizing the energy. For the asymmetric case, owing to the existence of parity symmetry, these parameters satisfy the relation:
\begin{equation}\label{parityrelation}
\left\{
\begin{aligned}
C^- & =  -C^+ = \frac{1}{\sqrt{2}}\\
\alpha^{-}_{\text{P},\text{A}} & =  \alpha^{+}_{\text{P},\text{A}} \\
\lambda^{-}_{\text{P},\text{A}} & =  -\lambda^{+}_{\text{P},\text{A}}
\end{aligned}
\right.
\end{equation}
With Eq. (\ref{anatz2}) and Eq. (\ref{parityrelation}) one can recover the result of the asymmetric Rabi model in Ref. \cite{variation}. For the asymmetric case, the symmetry between the up- and down-spin directions is broken and the convenience provided by the parameter constraint Eq. (\ref{parityrelation}) is no longer available. Thus, Eq. (\ref{anatz2}) implies a new variational scheme, which few references have mentioned.

\begin{figure}[t]
\includegraphics[width=0.8\columnwidth]{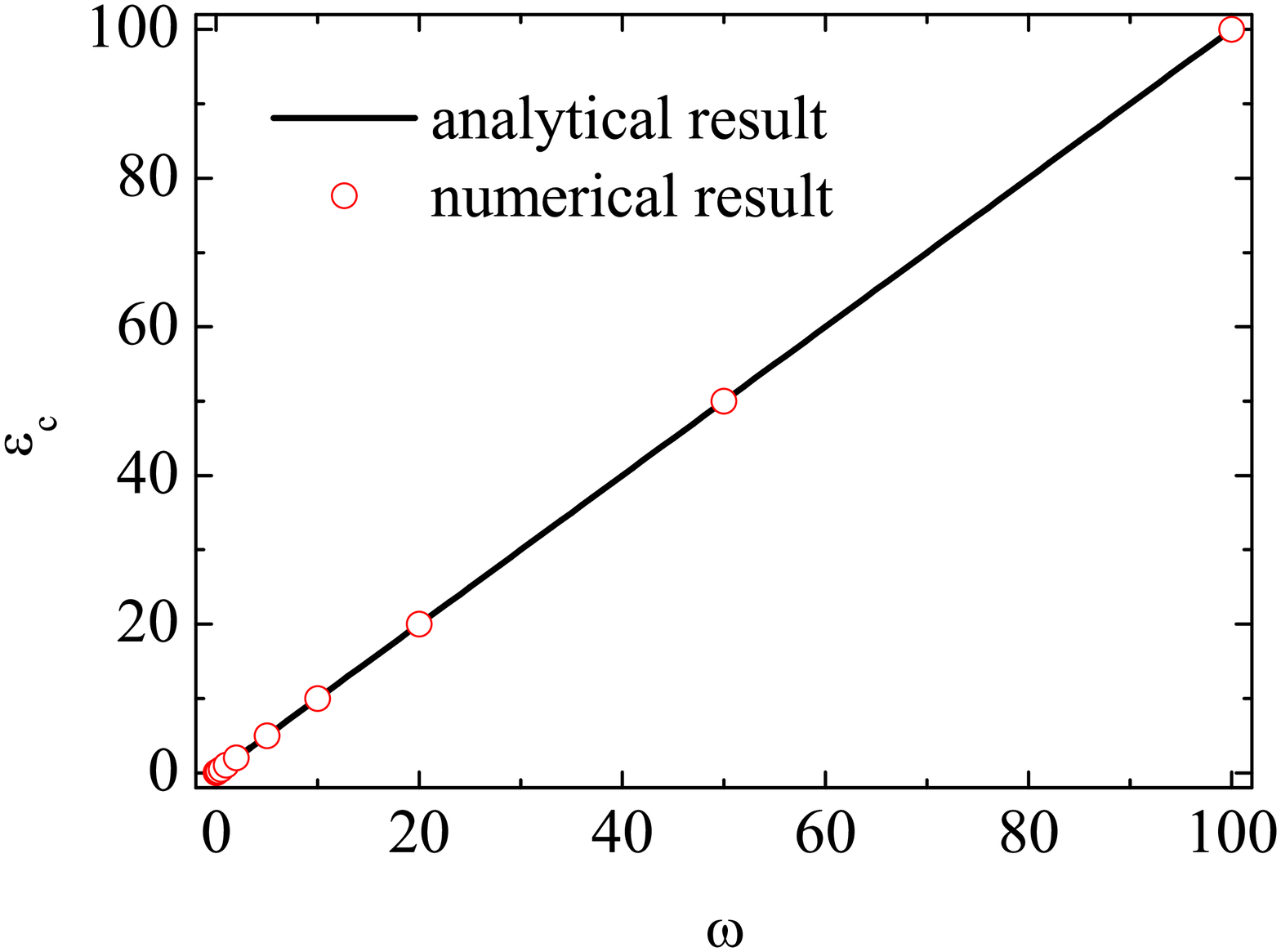}
\caption{(Color online) Critical asymmetry strength $\epsilon_c$ as a function of oscillator frequency $\omega$. The black solid line indicates the relation $\epsilon_c=\omega$ derived by our variational method based on an analytical calculation. The red circles are extracted by the numerical results of the small $g$ displacement.}
\label{ec}
\end{figure}
Now, let us return to the analytic expression for $\epsilon_c$. We take the trial wave function (\ref{anatz2}) as a starting point to discuss the relation between the critical asymmetry strength and the harmonic frequency. Since $\epsilon_c$ marks the asymmetric strength where $g_0$ approaches to zero, it is sufficient limit our discussion to the weak coupling limit, namely, $g \rightarrow 0$. In this case, the polaron and anti-polaron do not separate and show as a single wave-packet, which can be approximated by a single coherent state. Thus, the following variational form is sufficient:
\begin{equation}
|\Phi\rangle=\alpha_1|-_z\rangle|\lambda_1\rangle+\alpha_2|+_z\rangle|\lambda_2\rangle,
\end{equation}
where $\alpha_{1,2}$ are variational parameters satisfying the normalized condition:
\begin{equation}\label{normal}
\alpha_1^2+\alpha_2^2=1
\end{equation}
and $|\lambda\rangle=e^{\lambda(a-a^{\dag})}$ is the coherent state with variational parameter $\lambda$. It should be noted here that, unlike the trial wave function (\ref{trial}), the displacement amplitudes of the two coherent states are not the same.
Keeping only up to the second order of $\lambda_{1,2}$, the ground state energy is:
\begin{eqnarray}
&& E = \omega(\alpha_1^2\lambda_1^2+\alpha_2^2\lambda_2^2)+\Omega\alpha_1\alpha_2(1-\frac{1}{2}(\lambda_1-\lambda_2)^2) \nonumber\\
&& \hskip 1cm +2g(\alpha_2^2\lambda_2-\alpha_1^2\lambda_1)+\frac{\epsilon}{2}(\alpha_2^2-\alpha_1^2).
\end{eqnarray}
Minimization of the energy $E$ yields:
\begin{equation}\label{diff}
\frac{\partial E}{\partial \gamma }=0,
\end{equation}
where $\gamma$ takes $\alpha_1, \alpha_2, \lambda_1,$ and $\lambda_2$,  respectively.
A critical $\epsilon_c$ means that for $g=0$, $X_+=2\alpha_2^2\lambda_2$ should equal zero, so that $\lambda_2=0$.
The solution to Eqs. \eqref{diff}, subject to the normalized condition of Eq. \eqref{normal} and the critical point condition $\{\epsilon=\epsilon_c, \lambda_2=0\}$, eventually yields:
\begin{equation}
\left\{
\begin{aligned}
\lambda_1 & =  \frac{\sqrt{\Omega^2+\omega^2}-\omega}{\Omega^2}2g \\
\frac{\alpha_2}{\alpha_1} & =  -\frac{\sqrt{\Omega^2+\omega^2}-\omega}{\Omega} \\
\epsilon_c & = 2g\lambda_1+\omega\lambda_1^2-\frac{\Omega}{2} \left(\frac{\alpha_1}{\alpha_2}-\frac{\alpha_2}{\alpha_1} \right).
\end{aligned}
\right.
\end{equation}
It is seen that if $g \rightarrow 0$, the first two terms in $\epsilon_c=\omega$ vanish and the last two terms finally analytically yield $\epsilon_c=\omega$. Fig. \ref{ec} shows the relation between $\epsilon_c$ and $\omega$. Our numerical calculation and analytical results coincide very well. So far, we have provided a complete overview of the behavior of the overweighted anti-polaron, in the entire parameter regime of the model.

As a final discussion, we notice that the critical value of $\epsilon_c=\omega$ not only discriminates the different behaviors of the anti-polaron, but also shows its specialty in determining the structure of the energy spectra. In Ref. \cite{ec}, the authors found that the energy spectra avoid level crossing owing to the absence of parity symmetry in the asymmetry model, except for $\epsilon=N\omega$, where $N$ is an integer. For $N = 1$, the exceptional case of $\epsilon$ coincides with the critical asymmetry strength we obtained. This relation inspires us to speculate that $\epsilon_c$ is an intrinsic critical value in the model, which is not only crucial for the behavior of the polaron, but also may affect the some other excitation-related properties of the system.

\section{Summary}\label{section_con}
By numerically solving the ground state wave function, we
have systematically analyzed the behaviors of the
polaron and anti-polaron in the asymmetric Rabi model. Based on the competition and interplay among the four energy scales
involved in the model, we have extracted the basic physics:
(i) The competition between the tunneling and coupling
strength determines whether the polaron and anti-polaron
separate, which manifests as a phase-transition-like behavior around
$g_c$ in the small harmonic frequency limit. (ii) The
separation between the polaron and anti-polaron leads to a
dramatic suppression of the tunneling rate. For the small finite
asymmetry strength, the interplay between the asymmetry strength and the reduced
effective tunneling rate brings about another
phase-transition-like behavior at $g_0$, denoting the
occurrence of an overweighted anti-polaron in the spin-up
component. This additional transition-like behavior is very
sharp in the small harmonic frequency limit and in a higher
harmonic frequency, although becoming a crossover, still remains
sharper than that at $g_c$. (iii) There is a critical value
of the asymmetric strength, $\epsilon_c$, concerning the existence of
the displacement sign transition at $g_0$. For an asymmetric strength smaller than the harmonic frequency, the
aforementioned sign transition of displacement will occur; otherwise,
for a larger asymmetric strength, the anti-polaron can be always overweighted and
there is no sign transition. (iv) The overweighted
anti-polaron can be detected by a displacement measurement
using a quantum optical method, such as balanced Homodyne
detection.

\begin{acknowledgments}
This work is partly supported by the programs for NSFC, PCSIRT
(Grant No. IRT16R35), the national program for basic research and
the Fundamental Research Funds for the Central Universities of
China. ZJY also acknowledges the financial support of the Future and
Emerging Technologies (FET) Programme within the Seventh Framework Programme for Research of the European Commission, under FET-Open Grant Number:618083 (CNTQC).
\end{acknowledgments}

\end{document}